# Evidence of Cosmic-Ray Excess from Local Giant Molecular Clouds

Vardan Baghmanyan[1,4], Giada Peron[2], Sabrina Casanova[1,2], Felix Aharonian[2,3,4], and Roberta Zanin[5]
[1] Institute of Nuclear Physics PAN, Radzikowskiego 152, 31-342 Kraków, Poland; vardan.baghmanyan@ifj.edu.pl
[2] Max-Planck-Institut für Kernphysik, P.O. Box 103980, D-69029 Heidelberg, Germany
[3] Dublin Institute for Advanced Studies, 31 Fitzwilliam Place, Dublin 2, Ireland
[4] High Energy Astrophysics Laboratory, RAU, 123 Hovsep Emin Street, Yerevan 0051, Armenia
[5] CTA observatory, Via Piero Gobetti 93/3, I-40129 Bologna, Italy

## Abstract

We report the analysis of the Fermi-Large Area Telescope data from six nearby giant molecular clouds (MCs) belonging to the Gould Belt and the Aquila Rift regions. The high statistical γ-ray spectra above 3 GeV well described by power laws make it possible to derive precise estimates of the cosmic-ray (CR) distribution in the MCs. The comparison of γ-ray spectra of Taurus, Orion A, and Orion B clouds with the model expected from Alpha Magnetic Spectrometer (AMS-02) CR measurements confirms these clouds as passive clouds, immersed in an AMS-02-like CR spectrum. A similar comparison of Aquila Rift, Rho Oph, and Cepheus spectra yields significant deviation in both spectral indices and absolute fluxes, which can imply an additional acceleration of CRs throughout the entire clouds. Besides, the theoretical modeling of the excess γ-ray spectrum of these clouds, assuming $\pi^0$-decay interaction of CRs in the cloud, gives a considerable amount of an enhanced CR energy density and it shows a significant deviation in spectral shapes compared to the average AMS-02 CR spectrum between 30 GeV and 10 TeV. We suggest that this variation in the CR spectrum of Cepheus could be accounted for by an efficient acceleration in the shocks of winds of OB associations, while in Rho Oph, similar acceleration can be provided by multiple T-Tauri stars populated in the whole cloud. In the case of Aquila Rift, the excess in absolute CR flux could be related to an additional acceleration of CRs by supernova remnants or propagation effects in the cloud.

*Unified Astronomy Thesaurus concepts:* Gamma-rays (637); Cosmic rays (329); Giant molecular clouds (653)

## 1. Introduction

The Galactic population of cosmic rays (CRs) includes ultrarelativistic particles up to the so-called *knee*, namely, up to energies of at least $10^{15}$ eV. While the spectrum of CRs has long been thought to be featureless up to the knee, recent high-precision measurements of the CR population close to the Earth have unveiled several new features at high energies, such as similar rigidity-dependent hardening of proton and helium spectra at rigidities above ∼200 GV (Ahn et al. 2010; Aguilar et al. 2015). These discoveries have challenged the current paradigm of CRs, which predicts supernova remnants (SNRs) to be the major contributors of CRs up to the knee. The current assumption is that the locally measured spectrum of CRs is representative of the CRs everywhere in the Galaxy, except in the immediate vicinity of CR accelerators, namely, within 100 pc. In fact, as a result of the long CR diffusion timescale, up to $10^7$ yr in the turbulent magnetic fields, the CR density in the Galaxy is believed to level up.

Galactic diffuse γ-ray emission, originating from the interactions of CRs with the matter in the Galaxy, has long been used as a tool to infer the CR density and spectrum in the Galaxy. While the average Galactic CR (GCR) background density or CR "sea" should roughly amount to 1 eV cm$^{-3}$ (the local CR energy density), several indications of differences in the GCR distribution have been collected by γ-ray experiments in the GeV-to-TeV energy range. The EGRET data on the diffuse γ-ray emission from the Galactic plane at low latitudes ($|b| < 2°$) showed evidence of a softer >4 GeV γ-ray spectrum in the outer Galaxy compared to the inner parts, which indicates a variation of CR spectrum with the Galactic radius (Hunter et al. 1997). The measurements of diffuse Galactic γ-ray emission by the Fermi-Large Area Telescope (LAT) also yields a softening of the CR spectrum with the galactocentric distance that was accompanied by a decrease of the CR density (Acero et al. 2016; Yang et al. 2016). Besides, HESS detected very-high-energy diffuse γ-ray emission from the Galactic Center Ridge, which was accounted for by a significantly harder CR spectrum compared to local CR measurements (Aharonian et al. 2006). All these results support the hypothesis that the CR spectrum is likely to vary in different parts of the Galaxy.

The study of CR properties in the Galaxy is especially effective through the γ-ray observations of giant molecular clouds (GMCs), which coincide with clear peaks in the gas distribution along the line of sight from a given direction in the Galaxy. Correspondingly, one expects a peak in the γ-ray emission produced by $\pi^0$-decay interaction of CRs with the high-density gas. Thus, GMCs can be considered as *barometers* of the CR density in localized regions in the Galaxy. Of particular interest are the MCs in the star formation region of the Gould Belt, which because of their convenient isolated position in the sky are ideal candidates for γ-ray studies with Fermi-LAT (Casanova et al. 2010; Ackermann et al. 2012; Yang et al. 2014; Neronov et al. 2017; Aharonian et al. 2020). Besides, these sources are located within a few hundred parsecs from us, which suggests that the CR distribution in these sources should be compatible with the local CR experimental measurements. Therefore, any possible deviation can be ascribed to propagation effects or additional particle acceleration in the clouds. In particular, as predicted in a seminal paper by Cesarsky & Montmerle (1983) and recently shown by Aharonian et al. (2019), high-energy γ-rays can be produced by particles in star-forming regions accelerated by strong stellar winds induced by massive stars of OB associations. Furthermore, even though T-Tauri stars have insufficient power release (Cesarsky & Montmerle 1983), significant contribution to the CR flux from these low-mass stars cannot be excluded due to their large number in some clouds.

In the past decade, MCs in the different parts of the Galaxy have been studied with Fermi-LAT by many authors. The analysis of the closest (within ∼0.3 kpc) MC complexes found no difference between the shape of the γ-ray spectrum in these clouds and the spectrum expected from local CR measurement, only indicating ∼20% deviation in the CR density, that was accounted for as a result of CR anisotropy or variation in supernova rate (Ackermann et al. 2012). The study of several massive star-forming clouds mostly not belonging to the Gould Belt region did not find any hint of deviation from the local CR spectrum near the Earth below 18 GeV (Remy et al. 2017). The detailed Fermi-LAT analysis of the high latitude Gould Belt MCs by Neronov et al. (2012) found no evidence of CR density variation, which later was confirmed by Yang et al. (2014), Neronov et al. (2017), and Aharonian et al. (2020) with some indication of deviation in spectral indices. The absolute fluxes below 10 GeV were comparable to the direct measurements of local CRs measured by the PAMELA experiment (Yang et al. 2014).

Recently, Aharonian et al. (2020) reported the results of Fermi-LAT γ-ray observations of MCs from different locations over the Galactic disk. According to this study, several clouds in the galactocentric 4–6 kpc ring show evidence of enhanced CR density, which was interpreted by a possible contribution from active CR accelerators near MCs or an increase of CR density toward the Galactic center (GC). An enhancement of γ-ray flux by AGILE and Fermi-LAT was also found at (l, b) = [213.9, −19.5] in the Orion region, which was explained as reacceleration of CRs in the interstellar medium (ISM) by diffusive shocks produced by the interaction of the k-Ori wind and the OB star formation region (Cardillo et al. 2019). These results show that testing the CR level inside MCs occupied by different types of massive and low-mass stars is of great importance for checking the efficiency of the acceleration mechanisms described above.

Assuming that the γ-ray emission from MCs is only produced by CR interaction with the matter inside the cloud the γ-ray flux can be written as follows:

$$F_\gamma(E_\gamma) \propto \xi_N A \int dE_p \frac{d\sigma}{dE_\gamma} N(E_p), \quad (1)$$

with the $\xi_N$ factor representing the contribution from heavy nuclei in the cloud and ISM and the $A = M_5/d_{\rm kpc}^2$ factor, where $M_5 = M/10^5 M_\odot$ and $d_{\rm kpc} = d/1$ kpc. Here, $M$ is the total diffuse mass of the cloud that we calculated using the Planck dust opacity map at 353 GHz following Equation (1) described Yang et al. (2014):

$$M_{\rm dust} = m_{\rm H} \tau_D / \left(\frac{\tau_D}{N_{\rm H}}\right)^{\rm ref} A_{\rm angular} d^2, \quad (2)$$

where $m_{\rm H}$ is hydrogen mass, $\tau_D$ is the dust optical depth, and $A_{\rm angular}$ represents the angular area of the cloud. The reference value of $(\tau_D/N_{\rm H})^{\rm ref}$, taken from Table 3 in Planck Collaboration et al. (2011), was used in the estimation of A factors with corresponding 14% systematic uncertainty at 353 GHz. Although, all MCs with A factors greater than ∼0.4 can be detected by Fermi-LAT (Aharonian et al. 2020), we considered only the clouds with higher A factors, which due to better statistics can provide precise measurements at high energies (see Table 1).

The Letter is structured as follows. We first introduce the observations of CRs from different parts of the Galaxy as well as the selection of the MCs used in this work. In Section 2 the Fermi-LAT data selection and analysis including the derivation of the spectral energy distributions (SEDs) are presented. In Section 3 we show the results of the Fermi-LAT observations, derivation of the CR spectra in each cloud, and comparison with the experimental measurements of CRs. A summary is given in Section 4.

**Table 1**
Parameters of the Clouds Analyzed in This Paper

| Name | $l$ (deg) | $b$ (deg) | $D$ (pc) | A ($10^5 M_\odot$ kpc$^{-2}$) | Angular Area (deg$^2$) |
|---|---|---|---|---|---|
| Aquila Rift | 24.14 | 12.48 | 225 | 16.02 | 104 |
| Taurus | 171.04 | −15.32 | 135 | 5.63 | 32 |
| Rho Oph | 354.34 | 16.82 | 125 | 3.98 | 24 |
| Orion A | 211.83 | −18.80 | 490 | 3.83 | 26 |
| Cepheus | 107.94 | 15.07 | 860 | 3.73 | 29 |
| Orion B | 205.14 | −13.69 | 490 | 2.89 | 14 |

**Note.** Galactic coordinates in the second and third columns correspond to the centers of the clouds templates. The distances listed in the fourth column are taken from Zucker et al. (2019), while A factors in the fifth column were estimated using the Planck opacity map.

## 2. Fermi-LAT Spectral Analysis

In this study, we selected six GMCs from the Dame et al. (1987) CO Survey with the highest A > 2 factors. For each MC we used Fermi-LAT ∼11.2 yr data collected between 2008 August 4 (MET 239557417) and 2019 November 1 (MET 596851205). The physical parameters of these GMCs are reported in Table 1.

The analysis was performed with the publicly available Python package Fermipy (Wood et al. 2017) using P8R2_SOURC_V6 instrument response function and the 4FGL catalog (Abdollahi et al. 2020) for the selection of point sources. During the analysis we selected only the photons belonging to the source event class, and to avoid contamination from γ-rays produced in the upper atmosphere, we excluded events with zenith angles lower than 90°. To reduce the contribution from Galactic diffuse emission and point sources on analysis of MCs, we used only the data above 3 GeV, where Fermi-LAT has a better point-spread function. It also made it possible to minimize the contribution from low-energy breaks in the γ-ray spectrum around several GeV (Neronov et al. 2017) and obtain a precisely measured high-energy component of the γ-ray spectrum.

To model the γ-ray emission of GMCs, for each cloud we created a template using the Planck dust opacity map at 353 GHz[6] selecting only more homogeneous and central regions corresponding to $>5 \times 10^{-5}$ opacity, where according to Planck Collaboration et al. (2011) CO has the main contribution in the opacity. This selection also allowed us to exclude the systematic uncertainties that could be present if more fragmented regions were selected.

For the estimation of $\pi^0$-decay γ-ray flux, we used the same map with a size larger than the region of interest (RoI), cutting the regions corresponding to the cloud. In addition we also included the contribution from inverse Compton (IC) scattering in the estimation of Galactic background using the $^{S}Y^{Z}10^{R}30^{T}150^{C}2$ template from GALPROP[7] (Vladimirov et al. 2011), as well as Fermi-LAT extragalactic diffuse emission (iso_P8R3_SOURCE_V2_v1.txt). In the modeling of γ-ray emission, we did not take into account a contribution from bremsstrahlung emission, since as reported in Gabici et al. (2007), it does not have a considerable contribution above 100 MeV because of low $e/p \sim 0.01$ ratio in MCs.

---
[6] http://pla.esac.esa.int

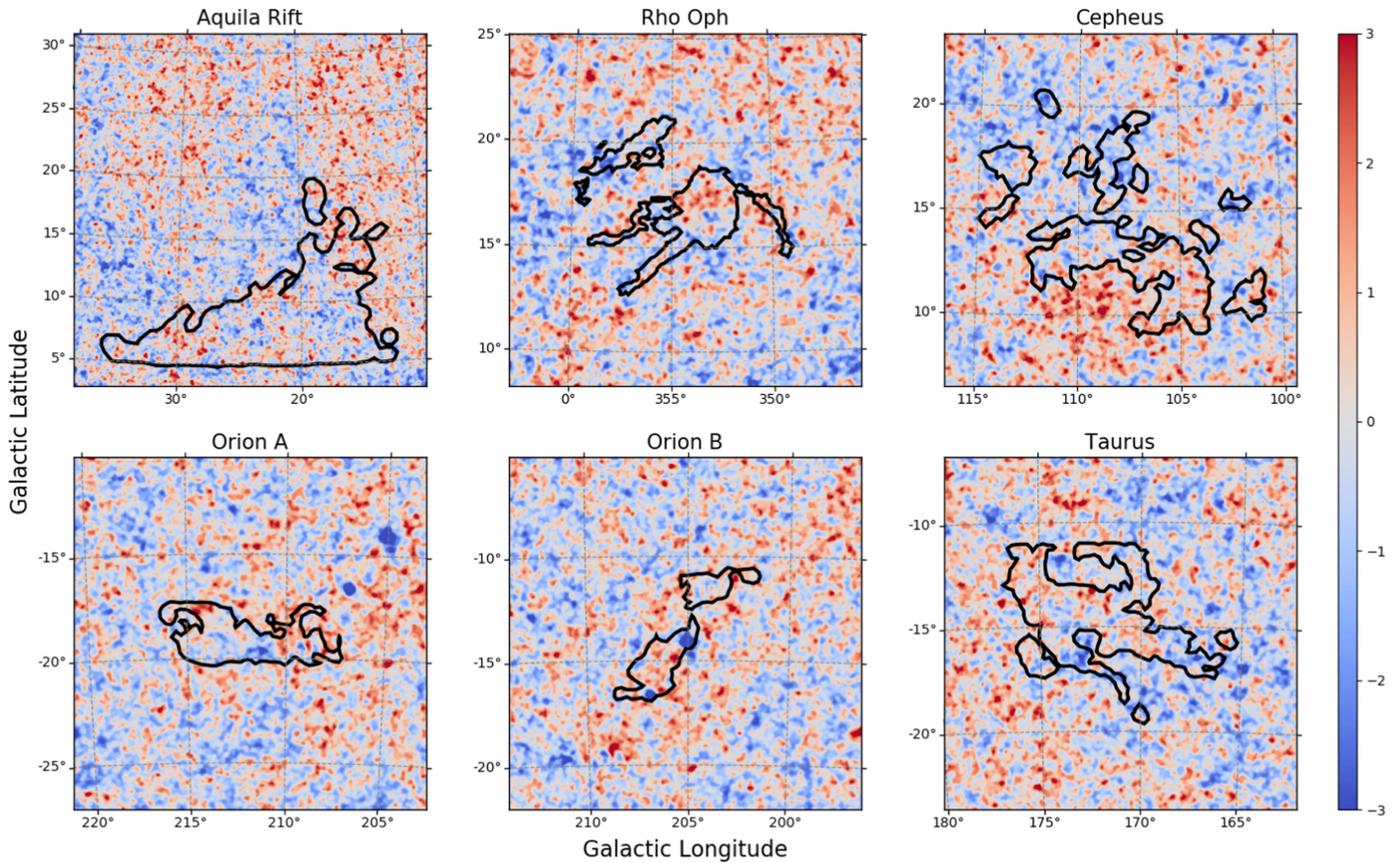

**Figure 1.** Residual maps of all RoIs. Black contours show the regions of the cloud templates with higher than $5 \times 10^{-5}$ dust opacity as used in the definition of the cloud models.

The data analysis for each individual GMC was done with a binned likelihood method using an RoI between 12° and 20° in radius depending on the size of the cloud. Since some of the clouds from our list are close to the Galactic plane we performed the spatial binning in a Galactic projection to minimize the contribution from the Galactic plane by using RoIs as far from the Galactic plane as possible. The RoIs for all clouds are shown in Figure 1 where black contours correspond to $>5 \times 10^{-5}$ opacity regions as used in the definition of the spatial extension of the clouds models. During the spectral analysis, the spectral parameters of extended sources as well as the normalization of point sources on or nearby the clouds were left as free parameters. In order to decrease the statistical uncertainties, we removed sources with the TS < 1 and re-ran the analysis. Despite the fact that our analysis is based on the latest 4FGL catalog and only the photons with the energies >3 GeV were used, in some clouds, we detected excess from several regions with $>5\sigma$ detection significance, which were added to the analysis. The best-fit source parameters of all clouds are shown in Table 2.

To check the quality of the spectral fitting, for each cloud we created the residual maps that are shown in Figure 1, which are scaled from $-3\sigma$ to $3\sigma$. As can be seen from these maps, the residuals of all clouds are mainly within $\pm 3\sigma$ standard deviation showing slight $>3\sigma$ excess in some clouds, which we considered as possible sources of systematic errors on the spectral energy distribution (SED) as will be explained below. Furthermore, the residual map of Orion B has significant negative $<-3\sigma$ excess from two regions on the top of the cloud; therefore, the best-fit spectral parameters as well as the SED of this cloud are less reliable.

In order to calculate the SED of each cloud, we divided the whole 3 GeV–1 TeV energy range into 23 equally spaced logarithmic bins and performed a binned likelihood fit for each energy bin using a single power-law model. During the fit, we left the normalization of the cloud as a free parameter and fixed the index and all parameters of points sources in the RoI as well as background sources to the best-fit values of the whole energy range. The derived SEDs after re-binning are shown in Figure 2 in red.

Besides the systematic uncertainties of the Fermi-LAT effective area,[8] which we included in the SEDs, we also checked possible sources of systematic errors such as the uncertainties of our Galactic diffuse emission model and new sources that appeared in our RoI, particularly those that are on or nearby the

---

[7] https://galprop.stanford.edu

[8] https://fermi.gsfc.nasa.gov/ssc/data/analysis/scitools/Aeff_Systematics.html

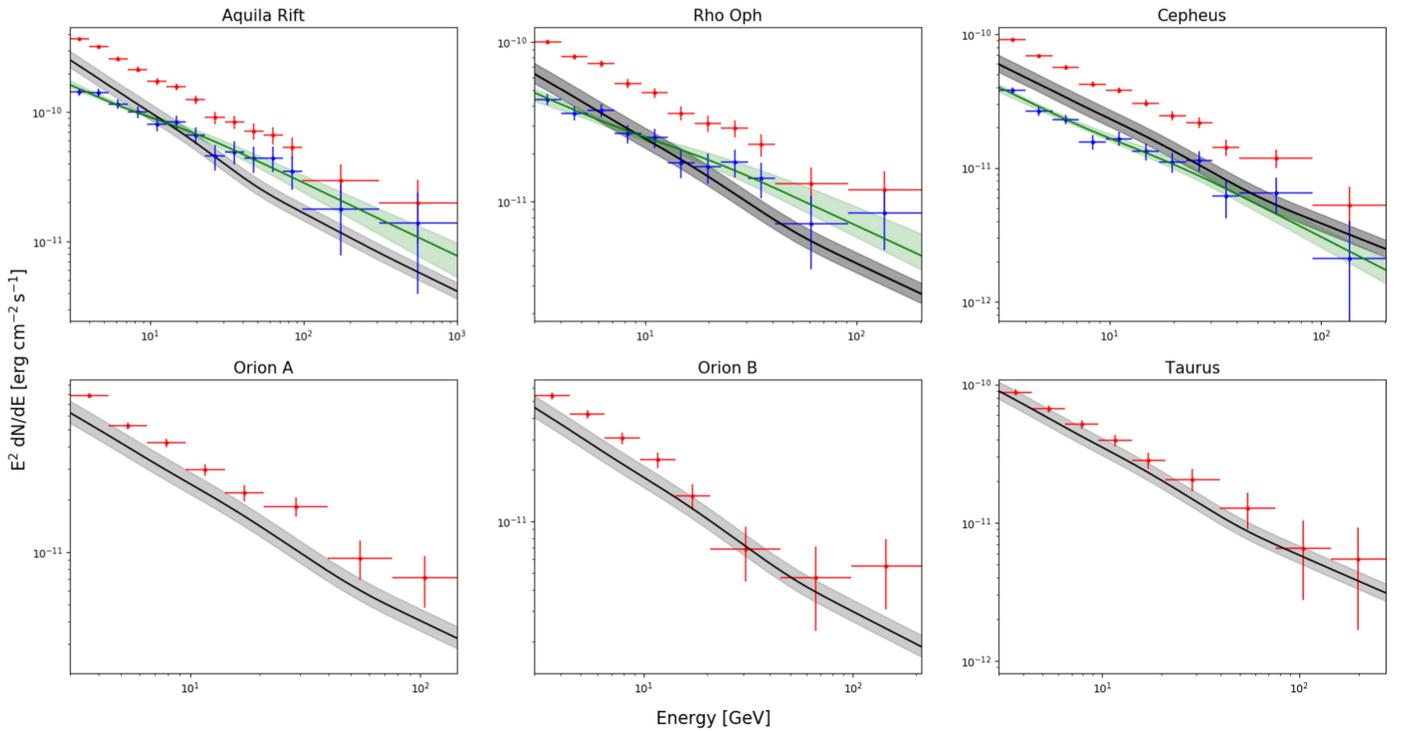

**Figure 2.** SEDs of three GMCs obtained from dust templates in comparison with the $\gamma$-ray spectrum derived from the direct CRs measurement by the AMS-02 experiment (black solid lines) where the shaded regions represent the 14% systematic uncertainties on the estimation of A factors. The green lines in the upper panels are the best-fit models of the $\gamma$-ray excess over the AMS-02-based $\gamma$-ray spectrum, where the shaded green regions correspond to $1\sigma$ uncertainties.

**Table 2**
The Best-fit Power-law Parameters of $\gamma$-Ray Spectral Analysis in Comparison with AMS-02-based $\gamma$-Ray Spectrum in the 3 GeV–1 TeV Energy Range

| Name | TS | Photon Index ($\Gamma$) | Flux at 3 GeV ($10^{-10} \times$ erg cm$^{-2}$ s$^{-1}$) | AMS-02-based Flux at 3 GeV ($10^{-10} \times$ erg cm$^{-2}$ s$^{-1}$) |
|---|---|---|---|---|
| Aquila Rift | 21547 | $2.64 \pm 0.01$ | $6.27 \pm 0.09$ | $3.11 \pm 0.46$ |
| Taurus | 8319 | $2.72 \pm 0.03$ | $1.34 \pm 0.04$ | $1.09 \pm 0.16$ |
| Rho Oph | 6381 | $2.67 \pm 0.03$ | $1.56 \pm 0.05$ | $0.77 \pm 0.11$ |
| Orion A | 9583 | $2.78 \pm 0.03$ | $1.07 \pm 0.03$ | $0.74 \pm 0.11$ |
| Cepheus | 6899 | $2.77 \pm 0.03$ | $1.24 \pm 0.03$ | $0.72 \pm 0.11$ |
| Orion B | 4899 | $2.84 \pm 0.04$ | $0.74 \pm 0.03$ | $0.56 \pm 0.08$ |

**Note.** The fluxes in fifth column are computed using the best-fit power-law normalization and index of AMS-02-based $\gamma$-ray spectrum for each MC in the energy range of 3 GeV–1 TeV, where errors correspond to 14% systematic uncertainties on A factors.

cloud. To explore systematic uncertainty coming from the selected Galactic diffuse and IC emission templates, we computed the SED after fixing the normalizations of these components to $(1 \pm 0.1)$ times the best-fit values of the global fit. In addition, we also produced the SED leaving the normalizations of these background diffuse components to vary in each energy bin. These tests showed no difference at the higher energies, while at lower energies below 10 GeV, it produced a minor influence on the SED that can be neglected. To check for possible systematic errors due to new sources on top or nearby the cloud, we produced the SEDs before and after adding these sources taking source detection thresholds of $3\sigma$, $4\sigma$, and $5\sigma$, although no significant differences between the SEDs were detected.

## 3. Results

To study the observed $\gamma$-ray flux we computed the expected $\gamma$-ray spectrum for each cloud based on the Alpha Magnetic Spectrometer (AMS-02) experimental proton data using the parameterization of $\pi^0$-decay cross sections by Kafexhiu et al. (2014). During this computation, we also included the energy-independent contribution of heavy nuclei both in the ISM and CRs with the factor of $\xi_N = 1.8$ as suggested by Mori (2009). The comparisons of observed and expected $\gamma$-ray spectra for each cloud are presented in Table 2 and Figure 2, where black shaded regions represent the 14% uncertainty on A factors coming from the systematic uncertainty on dust emissivity at 353 GHz that can be found in Table 2 of Planck Collaboration et al. (2011).

As can be seen from Figure 2, the $\gamma$-ray spectrum of Taurus coincides with the AMS-02-based $\gamma$-ray spectrum as already reported by several authors (Yang et al. 2014; Aharonian et al. 2020). However, in the case of Orion A and Orion B, there is a minor $\gamma$-ray excess as also reported by Yang et al. (2014) in the lower energies. However, taking into account the uncertainties no definitive conclusion on the $\gamma$-ray spectrum from Orion A and Orion B can be drawn.

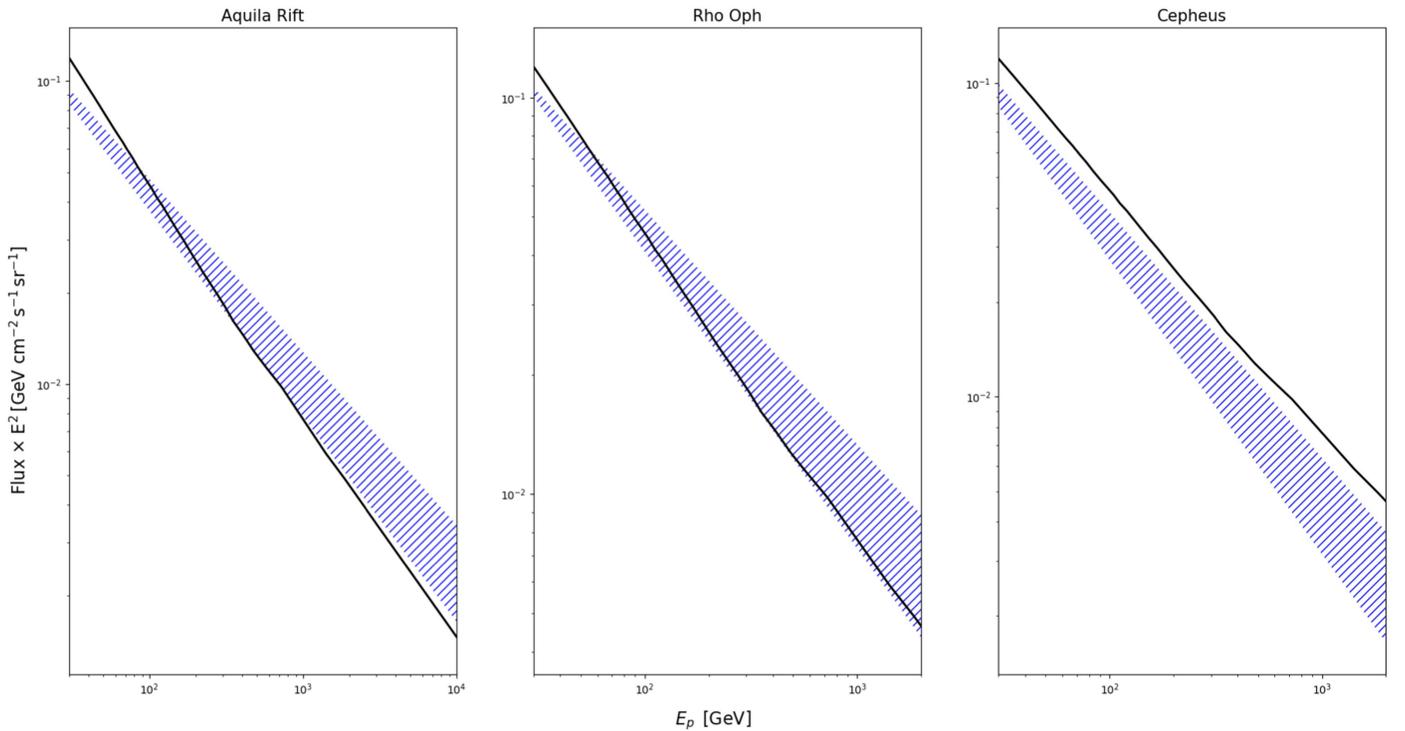

**Figure 3.** The enhanced CR spectra of three MCs obtained from fitting the excess of the γ-ray spectrum over the AMS-02-based γ-ray spectrum using the power-law model assuming the excess is produced by $\pi^0$-decay interactions inside the cloud. The shaded blue regions show both statistical errors in the fitted parameters and systematic uncertainties on A factors. The black curve represents the combination of AMS-02 experimental and analytical data calculated using Equation (3) of Aguilar et al. (2015) with the best-fit parameters.

The situation is different for Aquila Rift, Rho Oph, and Cepheus, which show clear enhancement of the γ-ray spectrum in the whole energy range, which is also obvious from the fluxes of observed and AMS-02 γ-ray photon fluxes integrated in the 3 GeV–1 TeV energy range. It is important to note here that the detected γ-ray excess in the above-mentioned MCs is most likely coming from the entire cloud instead of the local region(s) in the cloud. Evidence of this is the additional analysis of several small regions in these clouds that gave γ-ray spectra with a compatible excess over the AMS-02-based γ-ray spectrum as observed from the entire cloud.

Moreover, unlike Cepheus, which has pure power-law slope in agreement with the expected AMS γ-ray spectrum, the spectra of Aquila Rift and Rho Oph show a hint of slight hardening around 20 GeV that can be checked only with better statistics in higher energies.

In order to explore the origin of this enhancement in the γ-ray spectrum, for each cloud, we subtracted the AMS-02-based γ-ray spectrum from the observed one and fitted them assuming that the enhancement is produced by $\pi^0$-decay γ-rays. The subtracted spectra are shown in the upper panels of Figure 2 as blue solid lines where the uncertainties are the same as before the subtraction. The fitting was performed using naima v0.9.1 python package (Zabalza 2015), which uses the parameterization of $\pi^0$-decay cross sections by Kafexhiu et al. (2014) also taking into account the contribution from heavy nuclei. To obtain the CR energy distribution in the clouds, we applied the same method as in Aharonian et al. (2020) using a simple power-law model in the form of $dN/dE_p = N_0 (E/E_0)^{-\alpha}$, where $N_0$ corresponds to $\langle n_H = 1 \rangle$ cm$^{-3}$ and $d_{\rm kpc} = 1$, and therefore it should be normalized by an $\langle n_H \rangle = 1/d_{\rm kpc}^2$ factor. This approach allowed us to substitute the systematic uncertainties on $N_0$ into 14% uncertainty of the A factor in the computation of the CR spectrum:

**Table 3**
The Enhanced CR Power-law Spectral Indexes and CR Densities Calculated at 10 GeV

| Name | $\rho_{0,\,CR}$ $10^{-12} \times$ GeV$^{-1}$ cm$^{-3}$ | $\alpha$ |
| --- | --- | --- |
| Aquila Rift | 0.69 ± 0.28 | 2.62 ± 0.05 |
| Rho Oph | 0.86 ± 0.43 | 2.67 ± 0.08 |
| Cepheus | 0.91 ± 0.41 | 2.86 ± 0.08 |
| AMS-02 | 1.05 | 2.80 |

**Note.** For comparison, the average AMS-02 CR density at 10 GeV computed using the best-fit power-law normalization and index of experimental data between 30 GeV and 10 TeV is also presented. In the calculation of enhanced CR densities, both the statistical uncertainties on normalization estimated from the fit and 14% systematic uncertainty on A factors are taken into account.

$$\frac{dN}{dE_p} \propto \frac{N_0\, d_{\rm kpc}^2}{\langle n_H \rangle V} = \frac{m_p}{10^5 M_\odot A} N_0 = \rho_{0,\rm CR}. \quad (3)$$

The best-fitting parameters of the power-law CR distribution for each cloud are summarized in Table 3, while the corresponding enhanced CR spectra are shown in Figure 3. From this comparison, we can conclude that the enhanced CR spectra of Aquila Rift and Rho Oph are significantly harder than AMS-02 data, which in the 30 GeV–10 TeV energy range has $\alpha \sim 2.8$ power-law slope. In the case of Cepheus, the enhanced CR spectral index is similar to the AMS-02 spectrum showing only a slight softening.

## 4. Summary

Diffuse γ-ray emission in MCs above $10^9$ eV is mainly induced via $\pi^0$-decay production of CRs in the high-density gas in the cloud. Because of the proximity of MCs in the Gould Belt region, the study of γ-rays from MCs of this region allows us to precisely measure the distribution of CRs inside the cloud and check for possible deviations from local CR experimental measurements.

The γ-ray spectra of the MCs considered in this work are well described by a single power-law model above >3 GeV with spectral indexes ranging from 2.64 to 2.84. The spectra of Taurus, Orion A, and Orion B, shown in Figure 2, are well fitted with the model expected from AMS-02-based CR measurement, showing only slight deviations in normalization. This implies that in these clouds the γ-ray photons are mainly produced by hadronic $\pi^0$-decay interactions between the locally measured CR density and matter in the cloud. Unlike these sources, the spectra of Aquila Rift, Rho Oph, and Cepheus show evident excess over the AMS-02-based γ-ray spectrum. Besides the excess, the γ-ray spectra of Aquila Rift and Rho Oph also show a hint of slight hardening around 20 GeV that could be explained by the hardening of a proton or/and helium spectra at rigidities above ∼200 GV (Adriani et al. 2011; Aguilar et al. 2015).

The CR spectra of Aquila Rift and Rho Oph derived from the γ-ray excess over the AMS-02-based γ-ray spectrum show slightly hardening with respect to the AMS-02 spectrum with indices of 2.62 and 2.67, respectively, while the corresponding CR spectrum of Cepheus has a softer index with the value of 2.86. All of the above-mentioned arguments imply that in Aquila Rift, Rho Oph, and Cepheus an additional acceleration of CRs throughout the entire cloud should exist. As predicted by Cesarsky & Montmerle (1983) and shown by Aharonian et al. (2019), a possible acceleration of CRs can be provided by diffusive shocks induced by stellar winds, which can be originated in massive stars due to the release of huge mechanical energy. This assumption can work for Cepheus, which contains several star-forming regions such as three nearby OB associations (Kun et al. 2008). A similar acceleration of CRs in Rho Oph can be provided by several hundred T-Tauri stars (Bontemps et al. 2001), which, despite the fact that they have insufficient energy release (∼$10^{32-33}$ erg s$^{-1}$ per star; Cesarsky & Montmerle 1983), a large number of similar stars can provide the CR excess. The situation is more complicated in the case of Aquila Rift. In spite of containing several star formation regions (Komesh et al. 2019), most of them are very close to the Galactic plane and were not included in the template used in our analysis. Therefore, the CR energy density excess in this cloud should have a different origin, such as additional acceleration from nearby accelerators like SNRs or due to propagation effects in the cloud.

## ORCID iDs

Vardan Baghmanyan  https://orcid.org/0000-0003-0477-1614
Giada Peron  https://orcid.org/0000-0003-3255-0077
Sabrina Casanova  https://orcid.org/0000-0002-6144-9122